%% file: em.tex
\documentclass{emulateapj}

\lefthead{van der Wel et al.}
\righthead{Star Formation Rates of $z=1$ Early-Type Galaxies}
\slugcomment{Accepted for publication in ApJ}

\begin{document}

\title{Low Star Formation Rates for $z=1$ Early-Type Galaxies in the
  Very Deep GOODS-MIPS Imaging: Implications for their
  Optical/Near-Infrared Spectral Energy
  Distributions\altaffilmark{1,2}}

\author{A.~van~der~Wel\altaffilmark{3}, M.~Franx\altaffilmark{4},
  G.D. Illingworth\altaffilmark{5} P.G.~van~Dokkum\altaffilmark{6}}

\altaffiltext{1}{Based on observations with the \textit{Hubble Space
    Telescope}, obtained at the Space Telescope Science Institute,
  which is operated by AURA, Inc., under NASA contract NAS 5-26555}
\altaffiltext{2}{This work is based in part on observations made with
  the \textit{Spitzer Space Telescope}, which is operated by the Jet
  Propulsion Laboratory, California Institute of Technology under NASA
  contract 1407.}
\altaffiltext{3}{Department of Physics and Astronomy,
Johns Hopkins University, 3400 North Charles Street, Baltimore, MD
21218; wel@pha.jhu.edu}
\altaffiltext{4}{Leiden Observatory, Leiden University, P.O.Box 9513, NL-2300 AA Leiden, Netherlands}
\altaffiltext{5}{University of California
Observatories/Lick Observatory, University of California, Santa Cruz,
CA 95064}
\altaffiltext{6}{Department of Astronomy, Yale
University, P.O. Box 208101, New Haven, CT 06520-8101}

\begin{abstract}
  We measure the obscured star formation in $z\sim 1$ early-type
  galaxies.  This constrains the influence of star formation on their
  optical/near-IR colors, which, we found, are redder than predicted
  by the model by Bruzual \& Charlot (2003).  From deep ACS imaging we
  construct a sample of 95 morphologically selected early-type
  galaxies in the HDF-N and CDF-S with spectroscopic redshifts in the
  range $0.85<z<1.15$.  We measure their $24~\mu\rm{m}$ fluxes from
  the deep GOODS-MIPS imaging and derive the IR luminosities and star
  formation rates. The fraction of galaxies with $>2\sigma$ detections
  ($\sim 25~\mu\rm{Jy}$) is $17_{-4}^{+9}\%$. Of the 15 galaxies with
  significant detections at least six have an AGN.  Stacking the MIPS
  images of the galaxies without significant detections and adding the
  detected galaxies without AGN we find an upper limit on the mean
  star formation rate (SFR) of $5.2\pm 3.0~M_{\odot}~\rm{yr^{-1}}$,
  and on the mean specific SFR of $4.6\pm
  2.2\times10^{-11}~\rm{yr}^{-1}$.  Under the assumption that the
  average SFR will decline at the same rate as the cosmic average, the
  \textit{in situ} growth in stellar mass of the early-type galaxy
  population is less than $14\pm 7\%$ between $z=1$ and the present.
  We show that the typically low IR luminosity and SFR imply that the
  effect of obscured star formation (or AGN) on their rest-frame
  optical/near-IR SEDs is negligible for $\sim 90\%$ of the galaxies
  in our sample.  Hence, their optical/near-IR colors are most likely
  dominated by evolved stellar populations.  This implies that the
  colors predicted by the Bruzual \& Charlot (2003) model for stellar
  populations with ages similar to those of $z\sim 1$ early-type
  galaxies ($\sim 1-3~\rm{Gyr}$) are most likely too blue, and that
  stellar masses of evolved, high-redshift galaxies can be
  overestimated by up to a factor of $\sim 2$.
\end{abstract}


\section{INTRODUCTION}

A convenient way to estimate stellar masses of galaxies is through
modeling their spectral energy distributions (SEDs) with stellar
population models
\citep[e.g.,][]{worthey94,vazdekis96,bruzual03,maraston05}.  This is
the most efficient method to quantify the stellar mass function at
high redshift
\citep[e.g.,][]{kauffmann04,drory04,forster04,rudnick06,borch06}.  The
robustness of this method relies on the reliability of the assumed
model parameters, such as the star formation history.  The correctness
of the models, given the star formation history and other parameters,
is an obvious requirement as well.

With the advent of the Infrared Array Camera \citep[IRAC,][]{fazio04}
on the \textit{Spitzer Space Telescope} \citep{werner04} the
rest-frame near-infrared (near-IR) has become a commonly used part of
the SED to infer stellar masses of $z\gtrsim 1$ galaxies.  However,
especially in the near-IR different stellar population models differ
from each other \citep{maraston05}, which indicates that there might
be systematic biases in such mass estimates.  In \citet{vanderwel06a}
we used IRAC imaging of a sample of early-type galaxies at $z\sim 1$
in the GOODS-South field with dynamically measured masses in order to
infer the evolution of the rest-frame near-IR $M/L$ of early-type
galaxies between $z=1$ and the present. We compared this with the
evolution of the rest-frame optical $M/L$ and the predictions from
several stellar population models.  We found that the near-IR $M/L$ of
the $z\sim 1$ galaxies are lower (or, the rest-frame $B-K$ colors
redder) than predicted by, for example, the model by \citet{bruzual03}
for a range of model parameters.  This indicates that stellar mass
estimates inferred from near-IR photometry suffer from significant
systematic uncertainties \citep{vanderwel06b}.

\begin{figure*}
\includegraphics[angle=180,width=18cm]{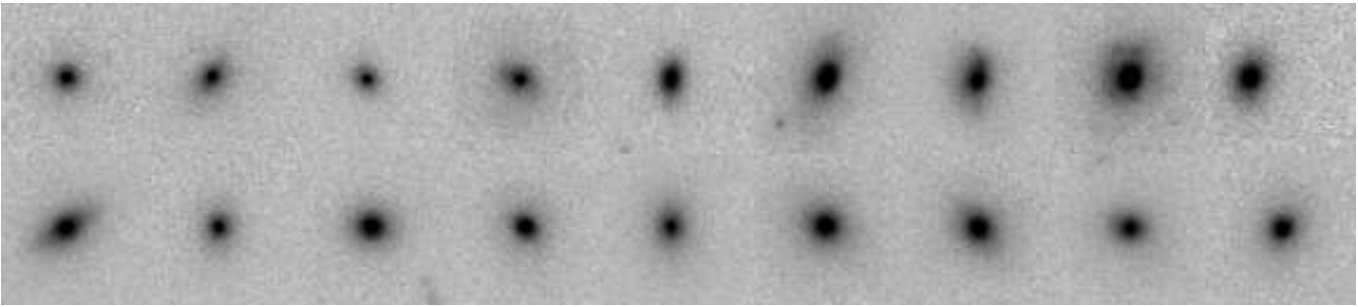}
\epsscale{0.9}
\caption{ ACS $z_{850}$-band cutouts of nine early-type galaxies with
bright ($>83~\mu\rm{Jy}$) $24~\mu\rm{m}$ detections (\textit{top row})
and nine randomly selected early-type galaxies from our sample with
similar rest-frame $B$-band luminosities but without significant
$24~\mu\rm{m}$ detections (\textit{bottom row}).  The cutouts are
$1\farcs 9$ on the side, which corresponds to $15~\rm{kpc}$ at $z=1$.
The morphologies of all 18 galaxies are S0/a or earlier ($T\leq 0$),
even though some of the galaxies with MIPS counterparts have somewhat
more irregular morphologies than the galaxies without MIPS
counterparts.}
\label{stamps}
\end{figure*}

The most straightforward interpretation of this result is that the
model colors are too blue, but this depends on the assumption that the
stellar populations of early-type galaxies are simple and can be well
described by a single burst stellar population.  If we add a
significant population of young, heavily obscured stars ($\gtrsim
10\%$ in mass), then we would reconcile the observations with the
Bruzual \& Charlot model \citep{vanderwel06b}.  Moreover, such a high
level of obscured star formation could account for the observed
increase of the stellar mass density of red sequence galaxies between
$z=1$ and the present \citep{bell04b,faber05,borch06,brown06}.  Hence,
determination of the star formation rate is interesting and relevant
in its own right.  Recently, \citet{rodighiero07} showed through
pan-chromatic SED modeling that, indeed, a significant fraction of the
early-type galaxy population at $z<1$ shows signs of hidden activity.

In this paper we construct a sample of morphologically selected
early-type galaxies with spectroscopic redshifts at $z\sim 1$, and
measure (upper limits of) their $24~\mu\rm{m}$ fluxes with the
Multiband Imaging Photometer for Spitzer \citep[MIPS,][]{rieke04} on
the \textit{Spitzer Space Telescope} (Sec. 2).  We use these to
constrain their star formation rates (SFRs) and the increase of their
stellar masses with cosmic time in Sec. 3.  Then we proceed, in
Sec. 4, to test the hypothesis that the rest-frame optical/near-IR
colors of $z\sim 1$ early types are significantly affected by obscured
star formation or Active Galactic Nuclei (AGN), and can explain the
observed red colors.  We summarize our conclusions in Sec. 5.
Throughout the paper we adopt the concordance cosmology,
$(\Omega_{\rm{M}},~\Omega_{\Lambda},~h) = (0.3,~0.7,~0.7)$.

\section{SAMPLE SELECTION AND MULTI-WAVELENGTH PHOTOMETRY}
We select galaxies with spectroscopic redshifts and early-type
morphologies from \textit{Hubble Space Telescope}/Advanced Camera for
Surveys (HST/ACS) imaging from the Great Observatories Origin Deep
Survey \citep[GOODS,][]{giavalisco04} in the Chandra Deep Field-South
(CDF-S) and the Hubble Deep Field-North (HDF-N).  Several
spectroscopic surveys conducted in the CDF-S
\citep{lefevre04,mignoli05,vanderwel05,vanzella06} are combined to
give 381 spectroscopic redshifts in the range $0.85< z< 1.15$.
Similarly, the compilation from \cite{wirth04} and the fundamental
plane study by \citet{treu05b} are combined to give 404 redshifts in
the same redshift range in the HDF-N.

Early-type galaxies are morphologically distinguished from late-type
galaxies by fitting Sersic models to the F850LP (hereafter, $z_{850}$)
GOODS images of all galaxies using the technique developed by
\citet{vandokkumfranx96}.  The best fitting $n$-parameter was
determined for every galaxy, with $n$ an integer, 1, 2, 3 or 4.  The
asymmetry $A$, the fraction of the total $z_{850}$ flux not situated
in the point-symmetric part of the light distribution, was also
determined for each object.  The criteria used to select early-type
galaxies are $n\geq 3$, $A<0.15$, $\chi^2<3$, and
$r_{\rm{eff}}>0\farcs 09$.  The latter three criteria are chosen upon
visual inspection of the images and the residuals of the fits.  The
limit $\chi^2<3$ excludes galaxies that are poorly fit by any model,
and the requirement $r_{\rm{eff}}>0\farcs 09$ excludes galaxies that
are very compact and for which the shape of the light distribution
cannot be reliably inferred.  The majority of the galaxies satisfying
these criteria have early-type morphologies as visually determined,
but a small number of Sa galaxies with clear spiral structure and
compact galaxies with small-scale irregularities, most likely mergers,
are rejected upon visual inspection.  The sample contains 95
early-type galaxies with E+S0+S0/a morphologies, with $T\leq 0$ in the
classification scheme of \citet{postman05}.  Total $z_{850}$-band
magnitudes are derived from the fitted profiles.  PSF-matched ACS and
GOODS IRAC images are used to measured rest-frame $B-K$ colors within
$5''$ diameter apertures, as described by \citet{vanderwel06a}.  Even
though the used spectroscopic surveys are neither complete nor
necessarily representative for the $z\sim 1$ galaxy population, there
is no reason to assume that our morphologically selected sample has a
bias in favor of or against early-type galaxies with obscured star
formation or AGN.  Therefore, we can assume that our sample is
representative for $z\sim 1$ early types as far as their IR properties
are concerned.

\begin{figure}
\epsscale{1.2}
\plotone{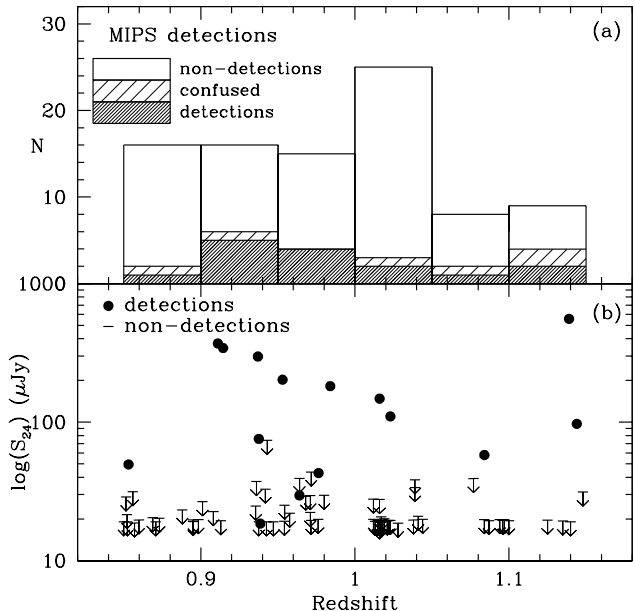}
\caption{ Panel (a): The number and redshift distribution of the
early-type galaxies in our sample.  The heavily shaded area indicates
the distribution of the galaxies with significant detections in the
$24~\mu\rm{m}$ MIPS image.  The non-shaded area indicates the
distribution of the galaxies without significant counterpart. For six
galaxies confusion prevented reliable identification of the MIPS
counterpart.  Panel (b): The redshift distribution of the
$24~\mu\rm{m}$ fluxes.  The galaxies with significant ($>2\sigma$) and
secure MIPS counterparts are indicated with the solid dots.  The other
data points are the $2\times \sigma$ fluxes for the non-detections,
i.e., galaxies with measured fluxes lower than $2\sigma$.  The six
possibly misidentified galaxies are omitted. }
\label{z_histM}
\end{figure}

We use the GOODS MIPS
images\footnote{http://data.spitzer.caltech.edu/popular/goods/Documents/goods\_dr3.html}
to obtain $24~\mu\rm{m}$ photometry for our sample of 95 early-type
galaxies.  Six galaxies turn out to be located in areas with exposure
times less than 10\% of the deepest parts of the images.  We
henceforth exclude these objects from the analysis, such that we have
a final sample of 89 early-type galaxies.  The publicly available
catalogs from the GOODS
team\footnote{http://www.stsci.edu/science/goods/} are used as a
reference to verify our own photometry, which we push deeper than than
the $83~\mu\rm{Jy}$ flux limit from the GOODS catalogs.  This limit is
a trade-off between completeness and the number of spurious
detections, but since we already know the positions of our objects,
deeper photometry is justified. Deeper photometry is necessary because
\citet{bell05} already concluded that the vast majority of early-type
galaxies at $z\sim 0.7$ are not detected down to $83~\mu\rm{Jy}$, an
upper limit that is too conservative for our goals.

We determined the two-dimensional background across the image with the
Sextractor software \citep{bertin96}, using the global background
setting.  We subtracted this background to obtain an image with
background set to zero.  Following, e.g., \citet{papovich04} and
\citet{zheng06}, we produce composite PSF-images for both fields,
comprised of isolated stars (identified in the ACS images), which we
use to create model images of MIPS sources at the positions of the
galaxies in our sample and objects in their vicinity.  The $3.6~\mu
\rm{m}$ IRAC images are used as a reference to identify these sources
and pinpoint their positions\footnote{The GOODS ACS, IRAC and MIPS images
are registered to the same world coordinate system with high accuracy,
with virtually no systematic offset ($<0.1"$) and a rms of $\sim
0.4''$ in the difference between the centroid positions of bright MIPS
sources and their IRAC counterparts, such that uncertainties therein
do not affect our measurements.}  The total flux of an object is
inferred from the PSF fitting model.  By default, the positions of the
$24~\mu m$ sources are left as free parameters, however, in case of
obviously erroneous results, we fix the positions at the IRAC
positions. This is necessary for very low $S/N$ sources.  In all
cases, IRAC and ACS images are used to visually identify the
counterpart of $24~\mu\rm{m}$ sources.  This procedure gives robust
$24~\mu\rm{m}$ flux measurements for 83 out of 89 galaxies in our
sample, of which 15 have a significant $24~\mu\rm{m}$ fluxes, with a
signal-to-noise ratio of at least two.  For six galaxies we cannot
produce reliable measurements because the centers of the low $S/N$
$24~\mu\rm{m}$ objects cannot be defined sufficiently accurate to
identify their counterparts with any confidence.  We label these
objects as 'possibly misidentified'.

The photometric error is dominated by noise and the uncertainty in the
background level.  In addition we include a 5\% error due to the
uncertainty in the aperture correction and a 2\% error due to the
uncertainty in the absolute photometric calibration.

10 of the galaxies with the brightest MIPS counterparts are also found
in the GOODS catalogs. The total fluxes as listed in the GOODS
catalogs agree within $\sim 5\%$ with the values that we derive,
except for one confused object for which we determined an accurate
flux measurement with the deconvolution method described above.  As an
\textit{a posteriori} verification of our morphological classification
methodology we show ACS $z_{850}$-band cutout images of nine of these
galaxies in Fig. \ref{stamps}, together with nine randomly selected
galaxies without significant MIPS counterparts: even the IR-bright
galaxies in our sample are genuine early types, as far as their
optical morphologies are concerned. We conclude that our morphological
selection criteria described above are sufficiently stringent to
exclude all late-type galaxies.

We list the measured fluxes in Table 1, and in Fig. \ref{z_histM}a we
show the redshift distribution of our sample.  The mean redshift is
$z=0.984$.  The shaded regions in Fig. \ref{z_histM}a show the
galaxies with MIPS counterparts.  The lightly shaded regions indicate
the six galaxies with possibly misidentified MIPS counterparts.  The
fraction of galaxies in our sample with MIPS counterparts (typically
$\gtrsim 25~\mu\rm{Jy}$) is $f=0.17_{-0.04}^{+0.09}$.  If we adopt the
brighter flux limit of $83~\mu\rm{Jy}$, the limit used for the GOODS
MIPS catalogs, we find $f=0.11\pm 0.03$.  In Fig. \ref{z_histM}b we
show the $24~\mu\rm{m}$ flux distribution.  The typical flux of the
objects with significant detections is $S_{24}\sim 100~\mu\rm{Jy}$,
and ranges from $\sim 25~\mu\rm{Jy}$ up to almost $1~\rm{mJy}$.  As
mentioned above, most galaxies in the sample have no significant
$24~\mu\rm{m}$ counterparts. For those galaxies the $2\sigma$ flux
levels, i.e., $2\times$ the photometric error, are shown.

\section{CONSTRAINTS ON THE STAR FORMATION RATE}
We use $S_{24}$ to constrain the bolometric infrared luminosity and
SFR.  \citet{papovich06} have shown that this is feasible with
reasonable accuracy.  First, given $S_{24}$, we compute $L_{12}=\nu
L_{\nu,12~\mu\rm{m}}$ for $z=1.0$ and $L_{15}=\nu L_{\nu,15~\mu\rm{m}}$
for $z=0.6$.  With the conversions from \citet{chary01} (Equations 4
and 5) we estimate the associated values for the bolometric infrared
luminosity $L_{\rm{IR}}$.  We introduce a $K$-correction by
interpolating between the values for $L_{\rm{IR}}$ inferred from
$L_{12}$ and $L_{15}$ to obtain $L_{\rm{IR}}$ at the observed redshift
$z$ of each object.  The $K$-correction is generally small (typically
10\%) since all redshifts are in the range $0.85<z<1.15$.  Finally,
assuming a Salpeter IMF, $L_{\rm{IR}}$ is converted into a SFR as
$(1.71\times10^{-10}~L_{\rm{IR}}/L_{\odot})~M_{\odot}~\rm{yr}^{-1}$
\citep{kennicutt98}.

The systematic uncertainties in the derived $L_{\rm{IR}}$ and SFR are
considerable. According to \citet{chary01} the uncertainty in the
transformation of $L_{12}$ into $L_{\rm{IR}}$ is of order 50\%.
Furthermore, as noted by \citet{papovich06}, the models by
\citet{dale02} yield $L_{\rm{IR}}$ that are lower by a factor of 2-3
for the most luminous objects ($L_{\rm{IR}}>10^{12.5}~L_{\odot}$).
For less luminous objects, like the objects in our sample, the
differences are smaller, therefore we adopt a systematic uncertainty
of 50\%.  Finally, the conversion of $L_{\rm{IR}}$ into SFR is
uncertain by about 30\%, such that the total uncertainty in the
derived SFR is a factor of two.

We have 15 galaxies with significant and secure detections, six of
which have X-ray counterparts \citep{alexander03}, with total X-ray
luminosities in the range
$L_{\rm{X}}=1-25\times10^{42}~\rm{erg~s^{-1}}$, which most certainly
means that these galaxies harbor type 2 AGN.  This is corroborated by
the fact that in all cases at least half of $L_{\rm{X}}$ is due to
hard X-rays.

Assuming that the $24~\mu\rm{m}$ flux of the other nine galaxies is
due to dust heated by star formation, we find SFRs ranging from 5 to
80 $M_{\odot}~\rm{yr^{-1}}$.  The IRAC colors of the two galaxies with
SFR exceeding 50 $M_{\odot}~\rm{yr^{-1}}$ are consistent with the
colors of a star-forming galaxy, and, moreover, in the rest-frame UV
F435W and F606W ACS filters these galaxies clearly show irregular
morphologies.  We stress, however, that, according to their rest-frame
optical morphologies, these galaxies are genuine early-type galaxies
with $\sim 90\%$ of the $z_{850}$-band flux accounted for by a smooth
De Vaucouleurs profile (see Fig. \ref{stamps}).

\begin{figure}[t]
\epsscale{1.1}
\plotone{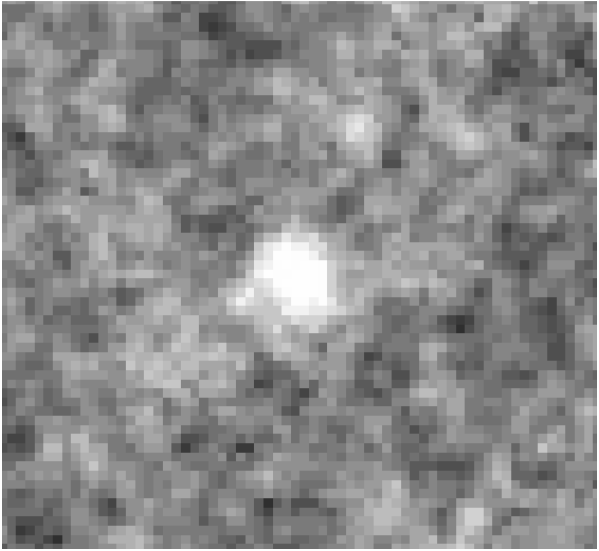}
\caption{ Stacked $24~\mu\rm{m}$ image (38" on the side) of the
  galaxies without significant individual detections and without X-ray
  counterparts.  The average flux is $6.8\pm 1.2~\mu\rm{Jy}$, which, at
  the average redshift of $z=0.97$, corresponds to a SFR of
  $1.2~M_{\odot}~\rm{yr}^{-1}$.  This SFR is an upper limit to the
  real SFR, as some of the flux might be due to AGN activity or
  silicate emission. }
\label{stack}
\end{figure}

In order to constrain the SFR of the galaxies without significant
individual detections we stack their MIPS images.  We omit the
galaxies with X-ray counterparts. The stacked image shown in Figure
\ref{stack} is created by co-adding the individual images, clipping
the pixels (outside a 6'' radius aperture centered on the fiducial
position of the co-added object) at $2.3\sigma$ to mask surrounding
objects \citep[see also, e.g.,][]{burgarella06}. Then we proceed to
determine the total flux within a $12"$ diameter aperture.  To compute
the total flux, we subtract the background (the median in a concentric
annulus between $30"$ and $40"$) and multiply by the aperture
correction 1.697.  The uncertainty is determined by the noise
properties of the stacked image, the systematic uncertainties in the
zero point calibration and aperture correction, and an additional
uncertainty of 5\% due to sub-pixel variations in the source positions
of the individual images \citep{zheng06}.  The measured flux is
$6.8\pm 1.2~\mu\rm{Jy}$ per galaxy.  This corresponds to a SFR of
$1.2~M_{\odot}~\rm{yr}^{-1}$ for a galaxy at the average redshift of
the sample ($z=1$), which should be regarded an upper limit to the
true SFR. Some of the flux will be due low-luminosity AGN, and, in
addition, low levels of silicate emission can be expected from
circumstellar dust around Asymptotic Giant Branch stars \citep[see,
e.g.,][]{bressan98,piovan03,bressan06}.  It is beyond the scope of
this paper to constrain the various contributions of the measured
flux.

The average SFR of the galaxies without significant detections (those
used to create the stacked image) and the nine galaxies with
significant detections but without X-ray counterparts is
$5.2\pm3.0~M_{\odot}~\rm{yr}^{-1}$.  This is an order of magnitude
lower than the SFR of the typical MIPS source at $z\sim 1$
\citep{perez05}. Also, $\sim 80\%$ of the star formation is accounted
for by $\sim 10\%$ of the galaxies, which suggests either interlopers
or that such obscured star-formation events in early-type galaxies are
short lived.

Next, we estimate an upper limit on the specific SFR.  We assume that
$M/M_{\odot}=2 \times L_{\rm{B}}/L_{\odot,\rm{B}}$, which is the
typical $M/L$ as was determined by recent $z\sim 1$ fundamental plane
studies \citep{vanderwel05,treu05b}.  The inferred masses are
typically in the range of $5\times 10^{10} - 2\times 10^{11}
M_{\odot}$.  We find an upper limit for the average specific SFR of
$4.6\pm 2.2\times10^{-11}~\rm{yr}^{-1}$, that is, a growth in stellar
mass of 5\% per Gyr.  This is two orders of magnitude lower than the
specific SFR of a typical MIPS source at $z\sim 1$, which has a
stellar mass of $M/M_{\odot}=10^{10} M_{\odot}$ \citep{perez05}. We
can use the specific SFR to crudely constrain the growth in stellar
mass of the early-type galaxy population between $z\sim 1$ and the
present.  In the case that the SFR remains constant for the population
as a whole, i.e., that obscured bursts of star formation are as
prevalent in the local universe as they are at $z=1$, the stellar mass
would increase by $35\pm 17\%$.  It is quite unlikely that the average
SFR in early types has remained constant over the past 7.5 Gyr, as
many studies have shown that the SFR has decreased by an order of
magnitude since $z\sim 1$
\citep[e.g.,][]{madau96,wolf03,lefloch05,bell05,perez05,zheng06}.  If
we assume that the SFR declines exponentially and by a factor of 10
between $z=1$ and the present, the growth in stellar mass is $14\pm
7\%$.  We stress that these numbers are upper limits due to the
various other potential contributors to the measured IR flux.  Most
likely, the true \textit{in situ} growth of the stellar mass of the
early-type galaxy population is still lower.

These upper limits are consistent with the residual stellar mass
growth of $\sim 7\%$ estimated by \citet{gebhardt03} and the residual
star formation of $\sim 2~M_{\odot}~\rm{yr^{-1}}$ derived by
\citet{koo05} for early-type galaxies and bulges in the Groth Strip
Survey.  \citet{treu05b} find significant young stellar populations in
low-mass early-type galaxies ($M<10^{11}~M_{\odot}$) at redshifts
$0.2<z<1$ in the HDF-N, which suggests a considerable growth in
stellar mass ($20-40\%$) between $z=1.2$ and the present.  For more
massive galaxies, they find that the growth in stellar mass is
negligible.  The upper limits that we derive here are marginally
consistent with a mass increase of more than $20\%$, but then we have
to assume that all the observed $24~\mu\rm{m}$-flux is due to star
formation, which is probably unrealistic.  The specific SFRs for
galaxies more and less massive than $10^{11}~M_{\odot}$ (the median
mass) are $5.1\pm 2.5\times10^{-11}~\rm{yr}^{-1}$ and $3.3\pm
1.6\times10^{-11}~\rm{yr}^{-1}$, respectively.  This difference is not
statistically significant since the co-added fluxes only differ from
each other on the $1.2\sigma$ level.  Still, if anything, the specific
SFR of high-mass early types is higher than that of low-mass early
types, but, in addition to the low significance of the measurement, we
should keep in mind that high mass galaxies are more likely to have
AGN that might contribute to the $24~\mu\rm{m}$ flux.  To reconcile
these results with the large fraction of young stars in low-mass early
types \citep{treu05b}, these young stellar populations must have
formed in other galaxies that later became part of an early type, or
at a time when the galaxies had not yet attained their early-type
morphologies.

We conclude that the \textit{in situ} SFR of the early-type galaxy
population at $z\sim 1$ is low, and can only account for an increase
in the stellar mass density of early-type galaxies by $\lesssim 20\%$
between $z=1$ and the present.  Additional mergers and/or
morphological transformations of galaxies are required to explain the
observed increase in stellar mass density of red galaxies by a factor
of two \citep{bell04b,faber05,borch06,brown06}.

\section{THE EFFECT OF STAR FORMATION ON THE OPTICAL/NEAR-IR COLOR}
\begin{figure}
\epsscale{1.2}
\plotone{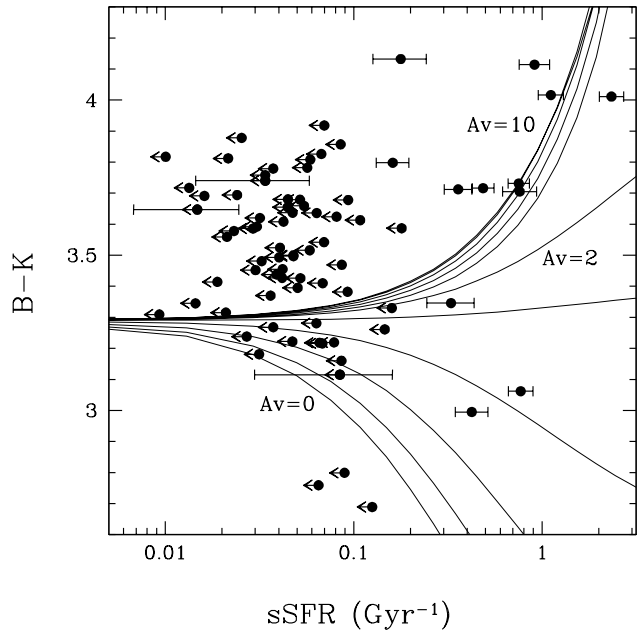}
\caption{ Specific SFR vs. rest-frame $B-K$ for galaxies with robust
  photometry (the six confused sources are left out).  Data points
  with error bars indicate galaxies with a significant MIPS detection,
  the other data points are upper limits for the galaxies without
  significant detections.  The lines are Bruzual \& Charlot models
  consisting of two components: an evolved stellar population and a
  young, obscured population with a constant SFR and $A_{\rm{V}}$
  increasing from bottom to top. See text for details.  The observed
  values of the specific SFR are much lower than predicted by the
  model for the majority of the galaxies, which implies that star
  formation likely does not significantly affect their $B-K$ colors.}
\label{ssfr_bk}
\end{figure}

Now we explore the question whether star formation affects the
rest-frame optical/near-IR colors of the galaxies in our sample, or
whether light from evolved stellar populations dominates that part of
their SEDs, as is usually assumed. In order to do so, we use the
(upper limits on the) specific star-formation rate derived in the
previous section, and compare this with the rest-frame $B-K$ color
(see Fig. \ref{ssfr_bk}).  The median $B-K$ color of the galaxies
without significant $24~\mu\rm{m}$ fluxes is $B-K=3.52$.  The median
$B-K$ of the 15 galaxies with significant $24~\mu\rm{m}$ fluxes is 0.2
mag redder, whereas the uncertainty in the $B-K$ color is only
$\lesssim 0.05$ mag.  This implies that obscured activity can indeed
affect the $B-K$ colors of galaxies.

We compare the colors of the galaxies in our sample with the expected
color of a $z=1$ galaxy with an evolved stellar populations. We
estimate this expected color as follows.  Given the measured evolution
of $M/L$ \citep{treu05b,vanderwel05}, the \citet{bruzual03} model
(with solar metallicity and a Salpeter IMF) predicts a certain amount
of evolution in $B-K$.  Therefore, from the $B-K$ color of local early
types \citep{vanderwel06a}, we can derive the expected color for
$z\sim 1$ early types. We find $B-K=3.29$.  We note that the $B-K$
colors of the galaxies with measured dynamical $M/L$
\citep{vanderwel06a} do not deviate from those of the larger sample
presented in this paper.

The median observed color is 0.2 mag redder than the expected color
$B-K=3.29$ (see Fig. \ref{ssfr_bk}). In particular, the galaxies
without significant $24~\mu\rm{m}$ counterparts are redder than
$B-K=3.29$. The question is how much room the upper limits on their
$24~\mu\rm{m}$ fluxes leave for attenuation of the $B-K$ color by
obscured star formation.  With the star-formation rates derived in
Sec. 3 we can constrain this scenario.

In Fig. \ref{ssfr_bk} we show two-component Bruzual-Charlot models,
consisting of an evolved stellar population with $B-K=3.29$ (see
above) and a 200 Myr old stellar population with a constant star
formation rate (solar metallicity, Salpeter IMF). Varying the age of
the young population with constant star formation between 50 and 500
Myr does not significantly change the models shows in
Fig. \ref{ssfr_bk}.  Besides the specific SFR, the attenuation
$A_{\rm{V}}$ (increasing from bottom to top) is the only other
variable. Only the young component is attenuated: $A_{\rm{V}}$ of the
evolved component with $B-K=3.29$ is assumed to be zero.  We assume
the \citet{calzetti00} extinction law.

As expected, models with low $A_{\rm{V}}$ predict blue colors for high
SFRs, whereas models with high $A_{\rm{V}}$ predict red colors.
Models with highly obscured star formation can reach $B-K$ colors that
match those of the galaxies in our sample. However, the associated
SFRs of those models are only observed for a handful of objects that
have significant MIPS detections. For the majority of the galaxies in
our sample, the observed SFRs are much lower than expected on the
basis of these models. This implies these models are inconsistent with
the red colors of the galaxies in our sample.  As indicated in Sec. 3,
the measured SFRs are systematically uncertain on the level of a
factor of two or so. The discrepancy with the model SFRs is much
larger than this (an order of magnitude for most galaxies).

Besides obscured SFR, obscured AGN could potentially also redden the
$B-K$ color of a galaxy. However, the vast majority of the galaxies in
our sample do not show evidence for nuclear activity in the IR or in
X-ray \citep[see also][]{rodighiero07}. It is highly unlikely that AGN
affect the $B-K$ colors of the majority of the galaxies without
leaving a trace at other wavelengths that are commonly used to
identify AGN.  The most straightforward conclusion is that the
\citet{bruzual03} model predicts colors that are too blue for stellar
populations in the age range of those of early-type galaxies at $z\sim
1$ ($1-3~\rm{Gyr}$, assuming solar metallicity).

\section{SUMMARY}
From MIPS $24~\mu\rm{m}$ imaging we derived constraints on the IR
luminosities of a sample of 89 morphologically selected early-type
galaxies at $z\sim 1$ with the purpose to identify obscured star
formation or AGN activity.  We find that 15 ($17^{+9}_{-4}\%$) have
$>2\sigma$ ($\sim 25~\mu\rm{Jy}$) $24~\mu\rm{m}$ counterparts.  The
X-ray luminosities of six of these indicate that an obscured AGN is
responsible for the IR emission. These findings are consistent with
recent work by \citet{rodighiero07} who study obscured activity in a
$z<1$ sample of morphologically selected early-type galaxies.

We derive an upper limit on the $24~\mu\rm{m}$ flux of the galaxies
without significant individual detections by stacking their MIPS
images.  When we add this sample to the galaxies with significant
detections but without luminous AGN we find $5.2\pm
3.0~M_{\odot}~\rm{yr^{-1}}$ as the upper limit of the average star
formation rate.  If we assume that the SFR of the early-type
population as a whole is constant between $z=1$ and the present, this
implies that the increase in stellar mass density of the early-type
galaxy population through \textit{in situ} star formation is at most
$35\pm 17\%$.  More realistically, if we assume that the average SFR
declines by an order of magnitude between $z=1$ and the present, i.e.,
if we assume that it follows the evolution of the average cosmic SFR,
we find $14\pm 7\%$.  This amount is too small to explain the observed
increase by a factor of $\sim 2$
\citep{bell04b,faber05,borch06,brown06}.

$10\%$ of the galaxies account for as much as $\sim 80\%$ of the
integrated IR luminosity, i.e., the majority of the galaxies are
quiescent in terms of star formation and AGN activity \citep[see
also,][]{bell05}.  The low IR luminosities of the galaxies imply that
the optical/near-IR colors of $z\sim 1$ early-type galaxies are
dominated by their evolved stellar populations, and are most likely
not significantly affected by obscured star formation or AGN (see
Sec. 4). Hence, the conclusions from \citet{vanderwel06a} and
\citet{vanderwel06b} hold: the rest-frame $B-K$ colors of the stellar
populations of $z\sim 1$ early-type galaxies are most likely redder
than predicted by the stellar population model from \citet{bruzual03},
and, consequently, stellar masses of evolved galaxies at high redshift
that are inferred from optical/near-IR photometry are overestimates by
up to a factor of two relative to stellar mass estimates at low
redshift.

\acknowledgements{ We thank the referee for very useful comments, and
we thank Eric Bell for stimulating discussions.  A. v. d. W. thanks
Andrew Zirm for providing MIPS photometry software, and acknowledges
support from NASA grant NAG5-7697.  }

\bibliographystyle{apj}

\input{tab1.tex}

\end{document}

%% file: tab1.tex
\begin{deluxetable}{lccccc}
\tabletypesize{\scriptsize}
\tablecolumns{6}
\tablewidth{0pt}
\tablenum{1}
\tablecaption{The Sample}
\tablehead {
\colhead{ID} &
\colhead{$S_{24}$} &
\colhead{$\log(L_{12})$} &
\colhead{$\log(L_{\rm{K}})$} &
\colhead{B-K} &
\colhead{z} \\
\colhead{} &
\colhead{$\mu$Jy} &
\colhead{$L_{\odot}$} &
\colhead{$L_{\odot}$} &
\colhead{} &
\colhead{}}
\startdata
    J123641.30+621618.4 &          $<17.9$	&  $<9.2$ &  9.8 & 3.47 & 0.85 \\
    J123604.28+621050.3 &          $<27.1$	&  $<9.4$ & 10.3 & 3.27 & 0.85 \\
    J123745.19+621655.8 &          $<20.2$	&  $<9.3$ & 10.2 & 3.92 & 0.85 \\
    J123648.61+621553.0 &          $<17.9$	&  $<9.2$ & 10.6 & 3.35 & 0.85 \\
    J123706.52+621818.7 &          $<29.5$	&  $<9.5$ & 10.2 & 3.86 & 0.86 \\
    J123714.48+621530.1 &          $<17.5$	&  $<9.2$ & 10.4 & 3.76 & 0.86 \\
    J123647.37+621628.4 &          $<17.9$	&  $<9.3$ & 10.5 & 3.41 & 0.87 \\
    J123714.24+621958.9 &          $<21.8$	&  $<9.4$ & 10.2 & 3.50 & 0.89 \\
    J123702.92+621428.1 &          $<18.5$	&  $<9.3$ &  9.4 & 2.69 & 0.90 \\
    J123646.13+621246.8 &          $<25.0$	&  $<9.5$ & 10.0 & 3.22 & 0.90 \\
    J123648.34+622010.2 &          $<21.3$	&  $<9.4$ & 10.3 & 3.68 & 0.91 \\
J123748.67+621313.2$^X$ &    $369.7\pm 11.2$	& $10.7$ & 10.5 & 4.02 & 0.91 \\
    J123601.81+621126.8 &          $<18.2$	&  $<9.3$ & 10.6 & 3.88 & 0.91 \\
    J123609.58+620845.1 &     $342.8\pm 9.7$	& $10.7$ & 10.7 & 3.72 & 0.91 \\
    J123728.95+621127.8 &          $<23.2$	&  $<9.5$ & 10.3 & 3.50 & 0.94 \\
    J123645.96+621101.3 &          $<35.0$	&  $<9.6$ & 10.2 & 3.61 & 0.94 \\
J123720.37+621523.9$^X$ &         \nodata 	&  \nodata & 10.6 & 3.78 & 0.94 \\
    J123712.70+621546.5 &    $297.5\pm 11.0$	& $10.6$ & 10.2 & 3.06 & 0.94 \\
    J123722.37+621543.7 &          $<17.9$	&  $<9.4$ & 10.3 & 3.48 & 0.94 \\
    J123734.36+622031.0 &    $75.7\pm 9.5$ 	& $10.1$ & 10.6 & 3.80 & 0.94 \\
    J123652.32+621537.5 &          $<17.8$	&  $<9.4$ & 10.4 & 3.74 & 0.94 \\
    J123654.28+621655.5 &          $<30.8$	&  $<9.6$ & 10.5 & 3.78 & 0.94 \\
J123640.15+621656.0$^X$ &          $<17.9$	&  $<9.4$ & 10.3 & 3.24 & 0.94 \\
    J123618.94+620844.8 &          $<17.9$	&  $<9.9$ & 10.6 & 3.69 & 0.94 \\
    J123600.63+621147.8 &          $<17.9$	&  $<9.4$ & 10.3 & 3.61 & 0.95 \\
J123630.05+620924.2$^X$ &   $202.4\pm 11.6$	& $10.5$ & 10.4 & 4.11 & 0.95 \\
    J123656.64+621220.5 &          $<18.0$	&  $<9.4$ & 10.3 & 3.49 & 0.95 \\
    J123652.55+620920.1 &          $<23.6$	&  $<9.5$ & 10.4 & 3.46 & 0.95 \\
    J123644.86+620844.7 &          $<27.7$	&  $<9.6$ & 10.2 & 3.22 & 0.97 \\
    J123745.15+621322.9 &          $<18.8$	&  $<9.5$ & 10.9 & 3.72 & 0.97 \\
    J123651.81+620900.2 &          $<18.3$	&  $<9.4$ & 10.2 & 3.40 & 0.97 \\
J123637.32+620831.2$^X$ &          $<41.0$	&  $<9.8$ & 10.2 & 3.16 & 0.97 \\
    J123620.68+620906.9 &          $<18.7$	&  $<9.5$ & 10.6 & 3.69 & 1.01 \\
J123650.30+622004.9$^X$ &          $<26.1$	&  $<9.6$ & 10.2 & 3.68 & 1.01 \\
J123644.39+621133.5$^X$ &          $<18.0$	&  $<9.4$ & 11.1 & 3.82 & 1.01 \\
    J123619.23+620923.2 &          $<18.1$	&  $<9.4$ & 10.5 & 3.31 & 1.01 \\
J123615.32+621135.0$^X$ &     $147.9\pm 6.0$	& $10.4$ & 10.5 & 3.71 & 1.02 \\
    J123620.28+620844.6 &          $<18.7$	&  $<9.6$ & 10.9 & 3.77 & 1.02 \\
    J123622.06+620851.0 &          $<17.0$	&  $<9.4$ & 10.5 & 3.59 & 1.02 \\
    J123630.69+620929.3 &          $<18.2$	&  $<9.4$ & 10.5 & 3.62 & 1.02 \\
    J123640.02+621207.7 &          $<17.9$	&  $<9.4$ & 10.3 & 3.18 & 1.02 \\
    J123658.09+621639.4 &          $<17.7$	&  $<9.4$ & 10.4 & 3.45 & 1.02 \\
    J123715.46+621212.2 &          $<18.1$	&  $<9.4$ & 10.4 & 3.68 & 1.02 \\
J123653.52+622018.9$^X$ &     $110.0\pm 7.1$	& $10.3$ & 10.0 & 3.71 & 1.02 \\
    J123556.74+621406.6 &          $<17.5$	&  $<9.4$ & 10.3 & 3.37 & 1.03 \\
    J123700.66+622103.2 &          $<31.6$	&  $<9.7$ & 10.5 & 3.43 & 1.04 \\
J123714.39+621221.5$^X$ &    $58.0\pm 9.0$	& $10.1$ & 10.0 & 3.35 & 1.08 \\
J123611.24+620903.4$^X$ &   $557.0\pm 11.9$	& $11.1$ & 10.6 & 4.01 & 1.14 \\
    J123613.50+621118.8 &          $<17.9$	&  $<9.6$ & 11.0 & 3.31 & 1.14 \\
    J123659.83+621934.5 &         \nodata 	&  \nodata & 10.3 & 3.54 & 1.14 \\
    J123727.86+622035.1 &          $<29.3$	&  $<9.8$ & 10.6 & 3.52 & 1.15 \\
    J033250.33-275246.8 &     $49.6\pm 9.4$	&  $9.7$ & 10.2 & 4.13 & 0.85 \\
    J033243.14-274242.0 &          $<18.4$	&  $<9.3$ & 10.3 & 5.44 & 0.86 \\
    J033238.36-274128.4 &          $<19.2$	&  $<9.3$ & 10.1 & 3.22 & 0.87 \\
    J033208.65-274501.8 &          $<19.0$	&  $<9.3$ & 10.5 & 3.56 & 0.87 \\
    J033235.45-274421.1 &         \nodata 	&  \nodata & 10.1 & 3.43 & 0.89 \\
    J033224.39-274624.3 &          $<18.5$	&  $<9.3$ & 10.3 & 3.65 & 0.90 \\
    J033232.04-274451.7 &          $<18.0$	&  $<9.3$ & 10.4 & 3.59 & 0.90 \\
    J033206.48-274403.6 &          $<20.6$	&  $<9.4$ & 10.3 & 3.83 & 0.96 \\
    J033223.26-275101.8 &          $<36.9$	&  $<9.7$ & 10.4 & 3.62 & 0.96 \\
    J033222.93-275434.3 &     $29.6\pm 9.1$	&  $9.4$ & 10.8 & 3.65 & 0.96 \\
    J033227.70-274043.7 &          $<27.4$	&  $<9.6$ & 10.4 & 3.66 & 0.97 \\
    J033256.92-274634.1 &          $<18.6$	&  $<9.4$ &  9.4 & 2.59 & 0.98 \\
    J033214.68-274337.1 &      $43.0\pm 9.2$	&  $9.4$ &  9.8 & 3.12 & 0.98 \\
    J033239.60-274909.6 &          $<27.8$	&  $<9.6$ & 10.8 & 3.58 & 0.98 \\
    J033254.23-274903.8 &     $181.9\pm 7.2$	& $10.7$ & 10.5 & 3.73 & 0.98 \\
\enddata
\end{deluxetable}
\clearpage
\begin{deluxetable}{lccccc}
\tabletypesize{\scriptsize}
\tablecolumns{6}
\tablewidth{0pt}
\tablenum{1}
\tablecaption{continued}
\tablehead {
\colhead{ID} &
\colhead{$S_{24}$} &
\colhead{$\log(L_{12})$} &
\colhead{$\log(L_{\rm{K}})$} &
\colhead{B-K} &
\colhead{z} \\
\colhead{} &
\colhead{$\mu$Jy} &
\colhead{$L_{\odot}$} &
\colhead{$L_{\odot}$} &
\colhead{} &
\colhead{}}
\startdata
    J033224.08-275214.6 &          $<18.1$	&  $<9.4$ &  9.7 & 3.33 & 1.01 \\
    J033229.93-275252.4 &          $<19.4$	&  $<9.5$ &  9.7 & 3.59 & 1.02 \\
    J033212.46-274729.2 &          $<18.4$	&  $<9.5$ & 10.3 & 3.44 & 1.02 \\
    J033217.77-274714.9 &         \nodata 	&  \nodata & 10.3 & 2.95 & 1.02 \\
    J033219.02-274242.7 &          $<18.9$	&  $<9.5$ & 10.1 & 3.54 & 1.02 \\
    J033210.12-274333.3 &          $<18.3$	&  $<9.5$ & 10.3 & 3.81 & 1.02 \\
    J033210.04-274333.1 &          $<18.3$	&  $<9.5$ & 10.7 & 3.81 & 1.02 \\
    J033244.29-275009.7 &          $<18.4$	&  $<9.5$ & 10.1 & 3.28 & 1.04 \\
    J033211.61-274554.2 &          $<36.0$	&  $<9.8$ & 10.5 & 3.64 & 1.04 \\
    J033217.91-274122.7 &          $<19.6$	&  $<9.5$ & 10.5 & 3.78 & 1.04 \\
    J033219.77-274204.0 &          $<18.7$	&  $<9.5$ &  9.7 & 3.26 & 1.04 \\
    J033231.37-275319.2 &          $<18.2$	&  $<9.6$ & 10.9 & 3.69 & 1.14 \\
    J033227.86-273858.2 &          $<36.7$	&  $<9.8$ & 10.3 & 3.38 & 1.08 \\
    J033301.27-275307.2 &          $<18.6$	&  $<9.5$ &  9.9 & 2.76 & 1.08 \\
    J033222.82-274518.4 &          $<18.3$	&  $<9.5$ &  9.8 & 2.80 & 1.09 \\
    J033216.17-275241.4 &          $<18.6$	&  $<9.5$ & 10.2 & 3.52 & 1.09 \\
J033219.30-275219.3$^X$ &          $<18.3$	&  $<9.5$ & 10.4 & 3.64 & 1.10 \\
    J033237.19-274608.1 &         \nodata 	&  \nodata & 10.9 & 3.60 & 1.10 \\
    J033231.22-274532.7 &          $<18.6$	&  $<9.5$ & 10.0 & 3.22 & 1.10 \\
    J033225.20-275009.4 &          $<18.3$	&  $<9.5$ & 10.1 & 3.41 & 1.10 \\
    J033245.15-274940.0 &         \nodata 	&  \nodata & 10.5 & 3.40 & 1.12 \\
    J033223.61-275306.3 &          $<18.4$	&  $<9.6$ & 10.4 & 3.43 & 1.12 \\
    J033218.52-275508.3 &     $97.3\pm 8.1$	& $10.4$ & 10.1 & 3.00 & 1.14 \\
\enddata
\tablecomments{ IDs, MIPS $24\mu\rm{m}$ fluxes, $12~\mu\rm{m}$
 luminosities, $K$-band luminosities, rest-frame $B-K$ colors and
 redshifts of the 89 galaxies in our final sample. $24\mu\rm{m}$
 $2\sigma$ values are given for objects with non-significant
 detections. The corresponding $12~\mu\rm{m}$ luminosities are
 calculated with that value. Objects with X-ray counterparts are
 labeled with $X$.}
\label{tab1}
\end{deluxetable}